\documentstyle{elsart}
\begin{document}
\begin{frontmatter}
\title{Growing interfaces in quenched disordered media}
\author[MdP]{L. A. Braunstein\thanksref{lbrauns}},
\author[MdP]{R. C. Buceta} and
\author[Murcia]{A. D\'{\i}az-S\'anchez}
\address[MdP]{Departamento de F\'{\i}sica, Facultad de Ciencias Exactas y 
Naturales, Universidad Nacional de Mar del Plata, Funes 3350, (7600) 
Mar del Plata, Argentina}
\address[Murcia]{Departamento de F\'{\i}sica, Universidad de Murcia, E-30071
Murcia, Espa\~na}
\thanks[lbrauns]{E-mail: lbrauns@mdp.edu.ar}
\begin{abstract}
We present the microscopic equation of growing interface with 
quench\-ed noise for the Tang and Leschhorn model [{\em Phys. Rev.} 
{\bf A 45}, R8309 (1992)]. 
The evolution equations for the mean heigth and the roughness are reached 
in a simple way. Also, an equation for the interface activity density 
({\em i.e.} interface density of free sites) as function of time is obtained.
The microscopic equation allows us to express these equations in 
two contributions: the diffusion and the substratum one.
All the equation shows the strong interplay between both contributions 
in the dynamics. 
A macroscopic evolution equation for the roughness is presented for this 
model for the critical pressure $p=0.461$. 
The dynamical exponent $\beta=0.629$ is analitically 
obtained in a simple way. Theoretical results are in excellent agreement 
with the Monte Carlo simulation. 
\end{abstract}
\end{frontmatter}
The growth of interfaces under nonequilibrium conditions is an interesting 
natural phenomenon. The interfaces has 
been characterized through scaling of the interface width $w$ with time $t$ 
and lateral size $L$. It is known that $w\sim L^\alpha$ for 
$t\gg L^{\alpha/\beta}$ and $w\sim t^\beta$ for $t\ll L^{\alpha/\beta}$. 
$\beta$ and $\alpha$ are called the dynamical and the roughness 
exponents, respectively.

The knowledge of the effects of the nonlinearities and the 
disorder of the media are important to describe the growing interface motion. 
Some experiments such as the growth of bacterial colonies and the motion 
of liquids in porous media, where the disorder is quenched, {\em i.e.} 
disorder due to the inhomogeneity of the media where the moving phase is 
propagating, are well described by directed percolation depinning (DPD) model 
proposed simultaneously by Tang and Leschhorn (TL) [1] and Buldyrev {\sl et.
al} [2]. In this work we will focus on TL model. 
This model predicts a dynamical exponent $\beta \simeq 0.63$ for $q_c=1-p_c=
0.539$, where $p_c$ is the critical pressure. However, some authors [2-4]
found a dynamical exponent $\beta$ faraway from the criticality.
Yang and Hu [4] found $\beta$ as function of $q$, analyzing the 
roughness as function of $t$. In a previous work, Braunstein and Buceta 
[5] show that the scaling law holds only at the criticality. 

We present now the Microscopic Equation (ME) for the TL model. The interface 
growth takes place in a lattice of edge $L$ with periodic boundary conditions. 
A random pinning force $g({\bf r})$ uniformly distributed in $[0,1]$ is assigned 
to every cell of the lattice.
For a given pressure $p$, the cells are divided in two groups, active (free)
cells with $g({\bf r})\le p$ and inactive (blocked) cells with $g({\bf r})>p$.
The interface is specified by a set of integer column heights $h_i$ 
($i=1,\dots,L$). During the growth, a column is selected at random with 
probability $1/L$ and compared his height with those of his neighbors. 
The time evolution equation for the interface in a time step $\delta t=1/L$ 
is
\begin{equation}
h_i(t+\delta t) = h_i(t)+\delta t\; [W_{i+1}+W_{i-1}+F_i(h'_i)\,W_i]
\;,\label{me}
\end{equation}
with
\begin{eqnarray*}
W_{i}&=&1-\Theta(h_i-\min(h_{i-1},h_{i+1})-2)\;,\nonumber\\
W_{i\pm 1}&=&\Theta(h_{i\pm 1}-\min(h_i,h_{i\pm2})-2)\{ [1-\Theta(h_i-h_{i\pm 2})]+\half
\delta_{h_i,h_{i\pm 2}}\} \;.\nonumber\\
\end{eqnarray*}
Here $h'_i=h_i+1$ and $\Theta(x)$ is the unit step function defined 
as $\Theta(x)=1$ for $x\ge 0$ and equals to $0$ otherwise. $F_i(h_i')$ 
equals to $1$ if the cell at the height $h_i'$ is free or active 
({\sl i.e.} the growth may occur at the next step) or $0$ if the cell is 
blocked or inactive. $F_i$ is the interface activity function.
Taking the limit $\delta t \to 0$ and averaging over the lattice we obtain 
($h=\langle h_i\rangle$)
\begin{equation}
\frac{dh}{dt}=
\langle 1-W_i \rangle + \langle F_i W_i\rangle\label{dhdt}\;.
\end{equation}
This equation allow us the identification of two separate contributions:
the difussion $\langle 1-W_i\rangle$ and the substratum 
one $\langle F_i\,W_i\rangle$.  
Yang and Hu [3] defined two kind of growth events: the event in 
which the growth occurs at the chosen site as a type $A$ (defined by us as 
local growth) and the event in which the growth occurs at the adjoint site 
as type $B$ (our growth by difussion). They count, in numerical simulation,
the events number $N_A(t)$ of type $A$ and $N_B(t)$ of type $B$ in a 
time interval $L$. They do not identify this terms as the substratum and 
the diffusion contributions to the mean height derivative. Remark that 
$N_A(t)\propto\langle F_i\,W_i\rangle$ and 
$N_B(t)\propto\langle 1-W_i\rangle$. They concluded that events $A$ roughen 
the interface while events $B$ flatten it. We shall see bellow that this last 
contribution (diffusion) enhances the roughness near the critical point in 
contradiction with the statement of Yang and Hu.  

We found an amazing numerical result: 
\begin{equation}
\langle F_i\,(1-W_i)\rangle=p\;\langle 1-W_i\rangle\;.
\end{equation}
We could not obtain analytically this result. Using this result, 
the interface activity density (IAD) $f=\langle F_i\rangle$ is
\begin{equation}
f=p\,\langle 1-W_i \rangle + \langle F_i\,W_i\rangle\;.
\end{equation}
The substratum contribution dominates the behavior of $f$ in
the earliest regime. 
Figure \ref{IAD} shows $f$ as function of time.
At the initial time, $f$ is equal to $p$ and in the earliest regime 
$f$ decreases as $t$ increases. In the intermediate regime 
($t\ll L^{\alpha/\beta}$) only at the critical value a temporal power law 
holds. 

From Eq.~(\ref{me}), the derivative of the square interface width 
$\d w^2/\d t$ (DSIW) can also be expressed by means of 
two additive contributions. The diffusion contribution is
\begin{equation}
2\,[\,\langle(1-W_i)\,\min(h_{i-1},h_{i+1})\rangle-
\langle 1-W_i\rangle\,\langle h_i\rangle\,]\label{dsiw_d}\;,
\end{equation}
and the substratum contribution is 
\begin{equation}
2\,[\,\langle h_i\,F_i\,W_i\rangle - 
\langle h_i\rangle\,\langle F_i\,W_i\rangle\,]\label{dsiw_s}\;.
\end{equation}
At short times, the diffusion process is 
unimportant because $W_i$ is close to one. As $t$ increases,
the behavior of this contribution depends on $q$. 
Notice, from Eq.~\ref{dsiw_d}, that the diffusion contribution may be either 
negative or positive. The negative contribution tends to smooth out the 
surface, this case dominated for small $q$.
The positive diffusion contribution enhances the roughness.  
This last effect is very important at the critical value. 
At this value, the substratum contribution is 
practically constant, but the diffusion contribution is very strong (see 
Figure~\ref{cp}),
enhancing the roughness in contradiction with the statement of Yang and Hu.
This last contribution has important duties on the power law behavior.
Notice that the roughness is the result of the competition between both
contributions.  
Generally speaking, the substratum roughen the interface while the diffusion 
flatten it for small $q$, but the diffusion also roughen the interface when $q$ 
increases. The diffusion is enhaced by local growth. The growth by diffusion 
may also increase the probability of local growth. This crossing interaction 
mechanism makes both, local growth and growth by diffusion, dominants in turn
near the criticality.

The ME approach to obtain the Langevin equation for the motion of surface 
has been used in the Das Sarma-Tamborenea model and the ballistic deposition 
with termal noise [6,7]. A perturbative expansion of the ME equation 
can be performed if one regularizes the step function. However, in the TL 
model it is not clear how to treat the quenched noise. The modeling of the 
quenched noise results an open question due to the complex nature of 
the growth. Nevertheless, we 
obtained a phenomenological equation for the roughness temporal derivative
[5]. The DSIW in the critical regime satisfy 
\begin{equation}
\frac{\d w^2}{\d t} =  c_1\,\exp[-2q(t+t_0)] + c_2\,(t+t_0)^{2\beta-1}\;,
\label{mac1}
\end{equation}
where the constants $c_j$ ($j=1,2$) and $t_0$ are related 
to the dynamics mechanism. $t_0$ is a small constant ($q\,t_0\ll 1$) that takes 
into account the initial condition $\d w^2/\d t\rfloor_{t=0}=p$.
Introducing the constant $c_1=2-1/\beta$,
and using the initial condition, at zero order in the exponential expansion 
around $qt_0=0$, we obtain $t_0=[(p-c_1)/c_2]^{1/(2\beta -1)}$.
The minimum of the DSIW is amazing reached at $t=\pi$, as show the 
Monte Carlo results (see Figure~\ref{th}). Using the second
derivative of interface width evaluated in the minimum
and replacing $c_1$, at zero order in $t_0$ expansion, we obtain
$c_2=(2q/\beta)\;\e^{-2q\pi}\;\pi^{-2(\beta-1)}$.
All the dynamic process is governed by the competition between the 
two terms of Eq.~(\ref{mac1}). Figure \ref{th} shows that at 
$t=1$ the contributions are equal. With this condition, 
we obtain the theoretical $\beta=0.629$ in excellent agreement with 
our numerical simulation results. The earliest regime is determinated
by the two terms until $t\simeq\pi$. Next the dynamics is governed by 
the power term.

Summarizing, we wrote the ME, for the evolution of the height, for the 
Tang and Leschhorn model. The ME allows us to derive the evolution of 
the mean height, the IAD and the DSIW in terms
of two separate contributions: the substratum and diffusion one. The
identification of this two separate contributions shows the crossing 
interaction between both. In particular, both are important at the 
criticallity for the roughness. We wrote a phenomenological macroscopic 
equation for the DSIW that allows us to predict the dynamical exponent
in a simple way. In the present, we are working on the DPD model of Buldyrev 
{\sl et. al}. Previous results show the same qualitative features for the 
diferents contributions of the ME and for the macroscopic equation as 
was expected for a model that belongs to the same universality class.

\begin{figure}
\vspace{60mm}
\caption{$f/p$ versus $t$. The parameter $q$ is
0.51 ($\Box$), 0.539 ($\bigcirc$) and 0.6 ($\bigtriangledown$). 
All cases shows the same behavior in the early time regime. 
The subcritical (supercritical) case shows that the IAD asymptotically 
goes to certain constant (zero). The critical case shows that the IAD goes 
as $t^{-\eta}$ (in solid line) with $\eta=0.400\pm10^{-3}$. 
\label{IAD}}
\end{figure}

\begin{figure}
\vspace{60mm}
\caption{Diffusion contribution to the DSIW (hollow symbols) and the 
substratum one (fill symbols) versus $\ln t$ for 
$q=0.539$ (circles) and $q=0.3$ (triangles).\label{cp}}
\end{figure}

\begin{figure}
\vspace{60mm}
\caption{For the critical $q=0.539$ we show the plot of the 
theorically DSIW versus time in solid line, Monte Carlo results in circles and
the first (second) term of DSIW versus time in dashed line {\bf a} ({\bf b}).  
The vertical solid line is used as a guide to show that both contributions 
are equal at $t=1$. The horizontal solid line is used as a guide to 
show the initial condition.\label{th}}
\end{figure}

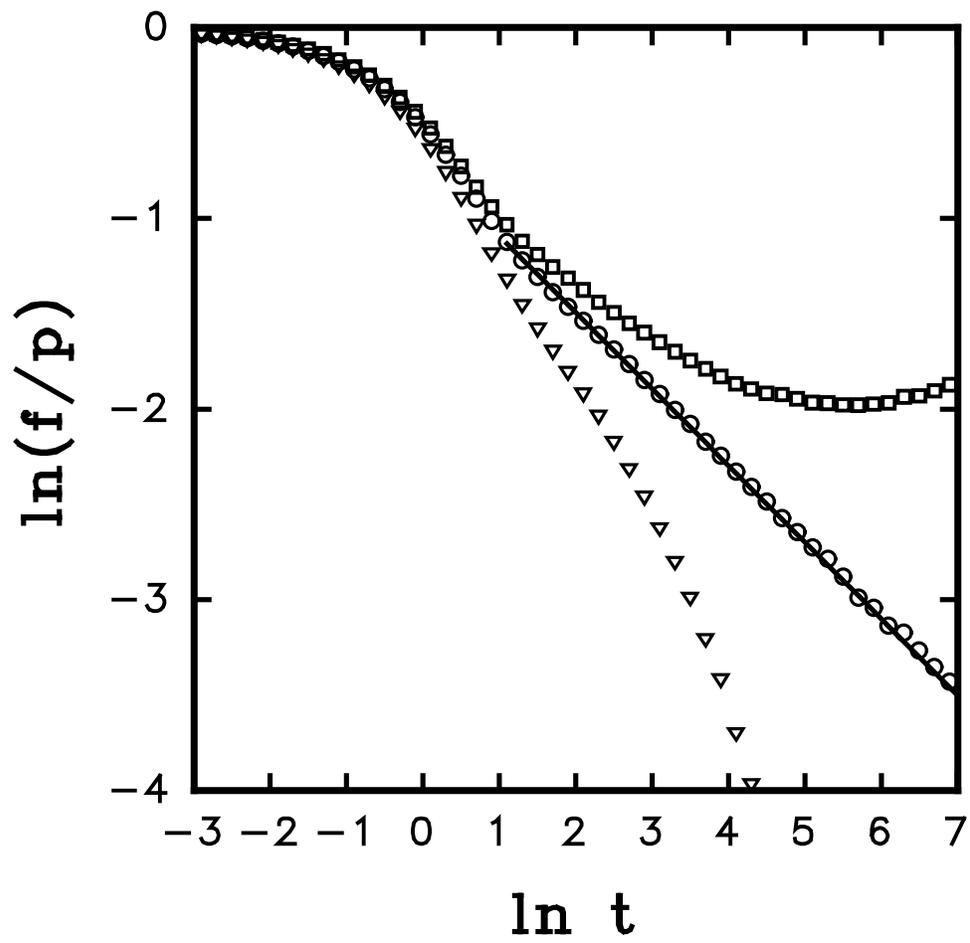
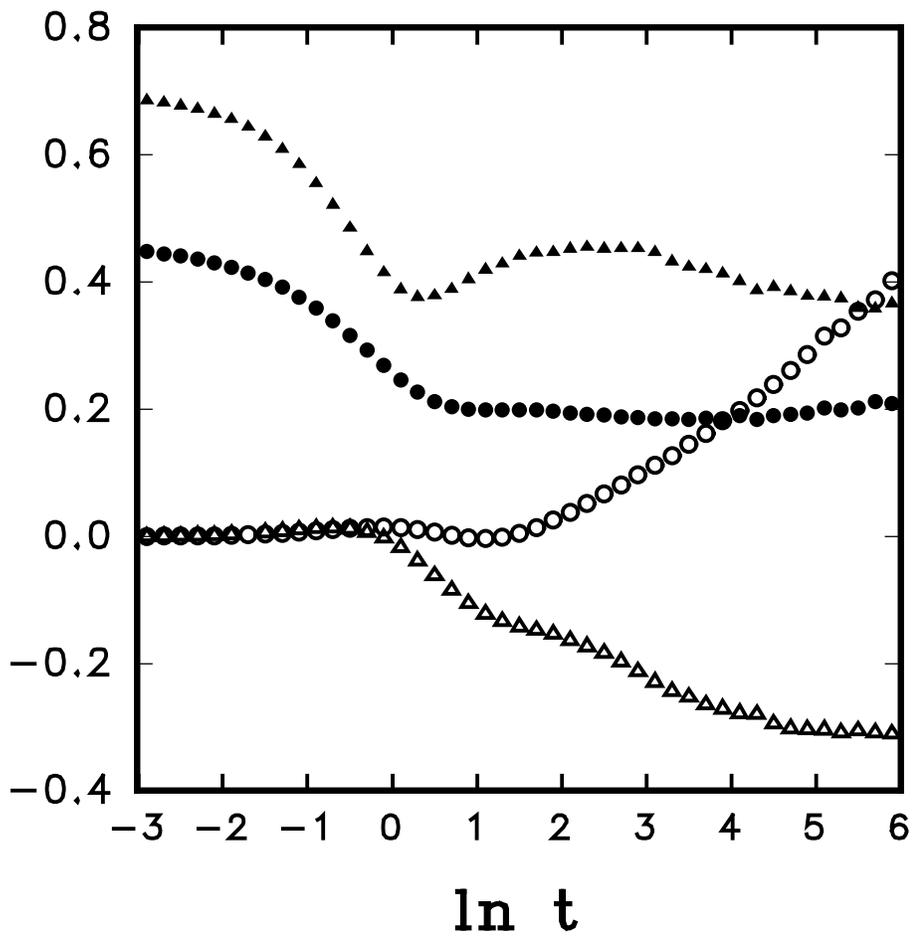
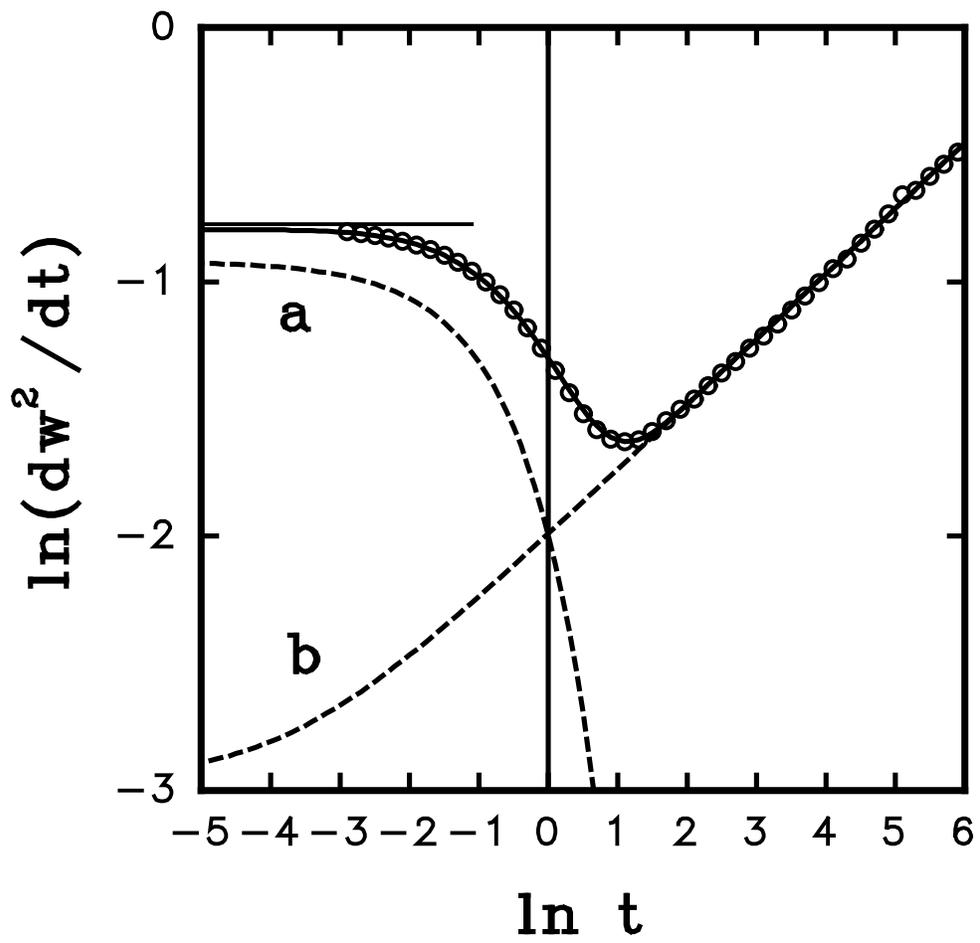
\end{document}